
%
%
\input phyzzx
\REF\Davis{R.~Davis, Gen. Rel. Grav. {\bf 19}, 331 (1987).}
\REF\Turok{N.~Turok, Phys. Rev. Lett. {\bf 63}, 2625 (1989).}
\REF\Vilenkin{M.~Barriola and
A.~Vilenkin, Phys. Rev. Lett. {\bf 63}, 341 (1989).}
\REF\Benneti{D.~Bennett and S.~Rhie, Phys. Rev. Lett. {\bf 65}, 1709
(1990).}
\REF\Paninal{L.~Perivolaropoulos, BROWN--HET--775 (November 1990).}
\REF\Bennetii{D.~Bennett and S.~Rhie, UCRL--IC--104061 (June 1990).}
 \REF\DGHR{
A. Dabholkar, G.W. Gibbons, J.A. Harvey,
F.R. Ruiz,
Nucl. Phys. \bf B340 \rm (1990) 33.} \REF\VAFA{B. Greene, A. Shapere,
C. Vafa and S.-T. Yau, Nucl. Phys. \bf B337 \rm (1990) 1.}
\REF\Rey{S.-J.~Rey, {\it Axionic String Instantons and Their
    Low-Energy Implications,} Invited Talk at Tuscaloosa Workshop on
    Particle Theory and Superstrings, ed. L.~Clavelli and B.~Harm, World
  Scientific Pub., (November, 1989); Phys. Rev. D {\bf 43}, 526 (1991).}
\REF\REYII{S.-J. Rey,
\sl On String Theory and Axionic Strings
and
Instantons, \rm talk given at `Particle \& Fields `91', Vancouver Canada
(1991); \sl Exact N=4 Superconformal
Field Theory of Axionic Instantons, \rm SLAC-PUB-5662 (1991).}
\REF\CHS{
 C. G. Callan, Jr. J. A. Harvey  and A. Strominger
Nucl. Phys. {\bf B359} (1991) 611.}
\REF\OT{
B. Ovrut and S. Thomas,
\sl Instantons in Antisymmetric Tensor Theories in Four-Dimensions,\rm
 UPR-0465T, (March 1991),
 and
Phys.Lett.{\bf B}267 (1991) 227.}
\REF\ANDY{A. Strominger,
Nucl. Phys.
\bf B343 \rm (1990) 167;
erratum \sl ibid \bf B353 \rm (1991) 565.
}
\REF\CHSII{
C. G. Callan, Jr. J. A. Harvey  and A. Strominger,
\sl Worldbrane Actions for String Solitons \rm , PUPT-1244(March 1991).}
\REF\Dine{M.~Dine and N.~Seiberg, Nucl. Phys. {\bf B301}, 357
                  (1988).}
\REF\Derend{J.~P.~Derendinger, L.~E.~Ib\'a\~nez, and
    H.~P.~Nilles, Phys. Lett. {\bf 155B}, 65 (1985); M.~Dine, R.~Rohm,
    N.~Seiberg, and E.~Witten, Phys. Lett. {\bf 156B}, 55 (1985).}
\REF\Ferrara{S.~Ferrara, D.~L\"ust, A.~Shapere, and S.~Theisen,
    Phys. Lett {\bf 225B}, 363 (1989).}
\REF\Cvetic{M.~Cveti\v c, A.~Font, L.~E.~Ib\'a\~nez, D.~L\"ust,
    and F.~Quevedo, Nucl. Phys. {\bf B361} (1991) 194.
    }
\REF\Kaplun{V.~Kaplunovsky, Nucl. Phys. {\bf B307}, 145 (1988).}
\REF\Dixon{L.~Dixon, V.~Kaplunovsky, and J.~Louis,
Nucl. Phys. {\bf B355}, 649 (1991); J.~Louis, SLAC--PUB--5527

(April 1991), in {\it PASCOS 1991 Proceedings}, P. Nath ed.,
World Scientific  1991.
}
\REF\Font{A.~Font, L.~E.~Ib\'a\~nez, D.~L\"ust, and F.~Quevedo,
    Phys. Lett. {\bf 245B}, 401 (1990);
          S.~Ferrara, N.~Magnoli, T.~R.~Taylor, and
    G.~Veneziano, Phys. Lett. {\bf 245B}, 409 (1990);
           P.~Binetruy and M.~K.~Gaillard, Phys. Lett. {\bf
    253B}, 119 (1991);
            H.~P.~Nilles and M.~Olechowski, Phys. Lett. {\bf
    248B}, 268 (1990).}
\REF\CV{M.~Cveti\v c, S.~J.~Rey and F.~Quevedo,
Phys. Rev. Lett. {\bf 63}, 565 (1991).}
\REF\CGR{M. Cveti\v c, S. Griffies and S.-J. Rey, {\it
Static Domain Walls in $N=1$ Supergravity},
UPR-474-T, YCTP-P43-91 (January 1992).}
\REF\Kim{J.~E.~Kim, Phys. Rev. Lett. {\bf 43}, 103 (1979);
    M.~Dine, W.~Fischler, and M.~Srednicki, Phys. Lett. {\bf 104B}, 199
    (1981); and Nucl. Phys. {\bf B189}, 575 (1981); M.~B.~Wise,
    H.~Georgi, and S.~L.~Glashow, Phys. Rev. Lett. {\bf 47}, 402 (1981);
    A good review is by J.~E.~Kim, Phys. Rep. {\bf 150}, 1 (1987).}
\REF\Sikivie{P.~Sikivie, Phys. Rev. Lett. {\bf 48}, 1156
    (1982);  G. Lazarides and Q. Shafi, Phys. Lett.{\bf 115B} (1982) 21;
    For a review, see A.~Vilenkin, Phys. Rep. {\bf 121}, 263
    (1985).}
\REF\modularform{B. Schoeneberg, \sl Elliptic Modular Functions, \rm
Springer, Berlin-Heidelberg (1970);
J. Lehner, \sl Discontinuous Groups and
Automorphic Functions, \rm
 ed. by the American Mathematical Society,
(1964).}
\REF\vafa{
P. Fendley, S. Mathur, C. Vafa and N.P. Warner, Phys. Lett.
\bf B243\rm (1990) 257.}
\REF\AT{ E. Abraham and P. Townsend, Nucl. Phys. {\bf B 351}, 313 (1991).}
\REF\WITTEN{
E. Witten, Comm. Math. Physics,
\bf 80 \rm (1981) 381.}
\REF\NESTER{
J.M. Nester, Phys. Lett. \bf 83A \rm (1981) 241.}
 \REF\GHW{
G.W. Gibbons, C.M. Hull, N.P. Warner,
Nucl. Phys. \bf B218 \rm (1983) 173 ;
C.M Hull, Nucl. Phys \bf B239 \rm (1984) 541.}.
\REF\Ipser{A. Vilenkin, Phys. Lett. \bf 133B \rm (1983) 177:
J. Ipser and P. Sikivie, Phys. Rev. \bf D30 \rm (1984) 712.}
\REF\COLEDEL{
S. Coleman
and F. DeLuccia, Phys. Rev. \bf D21 \rm (1980) 3305.}
\REF\CGRII{M. Cveti\v c, S. Griffies and S.-J. Rey,
{\it Nonperturbative Stability of Superstring Vacua}, UPR-494-T,
YCTP-P44-91  (February 1992).}
\REF\weinberg{S. Weinberg, Phys. Rev. Lett. {\bf 48}, 1776 (1981).}
\REF\Giveon{ A. Giveon, E. Rabinovici and G. Veneziano,
Nucl. Phys. {\bf B322}, 167 (1989).}
\REF\FONTII{A. Font {\it et al.}, in Ref.\Font\ .}
\REF\CD{M. Cveti\v c and   R. Davis,  to
appear as Univ. of Pennsylvania Preprint.}
\REF\LM{ D. L\" ust and C. Mu\~ nos, {\it Duality Invariant
Gaugino Condensation and One-Loop Corrected K\" ahler  Potentials
in String Theory}, CERN-TH 6358/91 (December 1991).}
\REF\CQRprep{M.Cveti\v c, F.~Quevedo, and S.-J.~Rey, unpublished;
M. Cveti\v c, UPR-485-T (September 1991), to appear
in the {\it Proceedings of the Strings 1991 Workshop}
(P. van Nieuwenhuizen), Stony Brook
May 20-25, 1991, World Scientific 1992.
}
\REF\CLO{M. Cveti\v c, J. Louis, and B. Ovrut,        Phys. Lett.
    {\bf 206B}, 227 (1988) and Phys. Rev. D {\bf 40}, 684 (1989).}
\REF\Gilm{R.~Gilmore, Lie Groups, Lie Algebras, and Some of Their
     Applications, John Wiley and Sons, (1974).}
\REF\Skyr{ G. Adkins, C. Nappi, and E. Witten, Nucl. Phys.
     {\bf B228}, 5521 (1983).}
\nopubblock \vskip0.2in
 \line{\hfill February 1992}
 \line{\hfill\caps UPR--496--T}
\def\hrf{\line{\hrulefill}}
\title{Stringy Domain Walls and Other Stringy Topological Defects
\foot{
Lectures
 Presented at Summer School on Particle Physics, Trieste, Italy, June 15
-- July 30, 1991}}
\author{Mirjam Cveti\v c}
\address{
Department of Physics\break
University of Pennsylvania\break
Philadelphia, PA 19104--6396\break}
\vfill
\singlespace
\vsize=630pt
\abstract{
We point out that the moduli sector of the $(2,2)$ string
compactification with its nonperturbatively preserved non-compact
         symmetries is a framework to study global topological
defects.
Based on the target space modular invariance of the
nonperturbative superpotential of the four-dimensional $N=1$
supersymmetric string vacua, topologically stable stringy domain walls
are found. Explicit supersymmetric solutions for the modulus field and
the metric, which
 saturate   the
Bogomol'nyi bound, are presented.  They interpolate between {\it
non-degenerate} vacua.  As a corollary, this defines
a new notion of vacuum degeneracy
of supersymmetric vacua. Nonsupersymmetric stringy domain walls are
 discussed as well.
The moduli sectors  with  more  than  one modulus and
the non-compact continous symmetry preserved
allow for     global monopole-type
and texture-type
configurations.}

\medskip
\line{\bf 1. Introduction\hfill}

Topological defects occur during the spontaneous break-down of gauge
symmetries, as a consequence of the nontrivial homotopy group $\Pi_n$ of
the vacuum manifolds.  Their existence has important cosmological
implications.  In particular global topological defects, like
textures\refmark{\Davis,\Turok}\ and more recently, global
monopoles\refmark{\Vilenkin,\Benneti}\ as well as global $\Pi_2$
textures\refmark{\Paninal,\Bennetii}\
were proposed as a source of large scale
structure formation.
On the other hand, in the framework of
 theories with extended gauge structures it
is often not aesthetically appealing  to impose
the existence of additional global (instead of local) gauge
symmetries, which would in turn allow for formation of global topological
defects.
 Here, we shall study
the moduli  sector of $(2,2)$ string compactifications which
provides a natural framework
for such global defects, with its potentially important physical
implications.

A new distinctive feature of superstring theories
is that
gravity and other moduli and matter fields
are on an equal footing,  so the effects of gravity can yield
distinctly new features. With the advent
of deeper understanding of semi-classical superstring theories
in a topologically nontrivial sector, various
stringy topological defects were discovered:
stringy cosmic strings
\refmark{
\DGHR\ , \VAFA },
axionic
instantons
\refmark{\Rey\ -\OT}
as well as
 related heterotic
five-branes and other solitons
\ \refmark {\ANDY\ , \CHS\ , \CHSII}
among others.

In this paper we will confine our attention to the
moduli sector of superstring vacua in four dimensions.
In $(2,2)$ string compactifications, where $(2,2)$ stands for $N=2$
left-moving as well as $N=2$ right-moving world-sheet supersymmetry,
there are massless fields -- moduli $T_i$ -- which have no potential,
 {\it i.e.}
$V(T_i)\equiv0$, to all orders in string loops\refmark{\Dine}\ .
Thus, perturbatively there is a large degeneracy of string vacua, since
any vacuum expectation value of moduli corresponds to the vacuum
solution.  On the other hand it is known that nonperturbative stringy
effects, like gaugino condensation\refmark{\Derend}\ and axionic string
instantons\refmark{\Rey}\  ,
give rise to the nonperturbative superpotential.

In the case of the modulus $T$ associated with the internal size of the
compactified space for the so-called flat background compactifications
({\it e.g.}, orbifolds, self-dual lattice constructions, fermionic
constructions) the generalized target space duality is characterized by
noncompact discrete group $PSL(2,{\bf Z})=SL(2,{\bf Z})/Z_2$ specified by
$$T\rightarrow{{aT-ib}\over{icT+d}}  \ \ \ ,
 ad-bc=1\ \ ,
 \{a, b, c, d\}\in {\bf Z}.$$
 If one assumes that the generalized target space duality is
preserved even
nonperturbatively\break
\refmark{\Ferrara,\Cvetic}\ ,
the form of
the nonperturbative superpotential is very
restrictive\refmark{\Cvetic}\ .
The fact that this is an exact symmetry
of string theory even at the level of nonperturbative effects is
supported by genus-one threshold
calculations\refmark{\Kaplun,\Dixon}\ ,
which in turn specify the form of the gaugino
condensate\refmark{\Font} .

This phenomenon has intriguing physical implications leading to the
stable supersymmetric domain walls\refmark{\CV ,\CGR}\ . This physics of
modulus $T$ is actually a generalization of the well known axion
physics\refmark{\Kim}\ introduced to solve the strong $CP$ problem in
QCD.  Spontaneously broken global $U(1)$ Peccei-Quinn symmetry is
non-linearly realized through a pseudo-Goldstone boson, the invisible
axion $\theta$.  Nonperturbative QCD effects through the axial anomaly
break explicitly $U(1)$ symmetry down to $Z_{N_f}$, by generating an
effective potential proportional to $1-\cos N_f\theta$.  This potential
leads to domain wall solutions \refmark{\Sikivie}\
with $N_f$ walls meeting at the axionic
strings\refmark{\Kim}\ .

The paper is organized as follows: In Chapter 2
we describe  global
supersymmetric domain walls as a warming up for the
stringy domain walls with gravity  included. In Chapter 3
local stringy domain walls are described, including
non-supersymmetric domain walls. In addition,
implications of  nonperturbative stability of supersymmetric
vacua are discussed.
Possiblity of other topological defects in  the
moduli sector of string theory is discussed    in Chapter 4.

\medskip
\line{\bf 2. Global Stringy Domain Walls\hfill}
\medskip

As an instructive example let's
first consider a global supersymmetric theory
with
$$
L = K_{T \bar T} |\nabla T|^2 + K^{T \bar T} |\partial_T W(T)|^2
\eqno (1)
$$
Here,
$ K_{T \bar T} \equiv \partial_T \partial_{\bar T} K(T, \bar T)\  $
is the positive definite metric on the complex modulus space and
the superpotential, $W$,
is a rational polynomial $ P(j(T))$
of the modular-invariant
function $j(T)$\refmark{\modularform}\ ,
{\it i.e.}
a modular invariant
form of $PSL(2,\bf Z)$.
The potential
$$V\equiv K^{T \bar T} |\partial_T W(T)|^2
= G^{T \bar T}
|\partial_j  P(j) \partial_T j(T)|^2$$
has at least
two isolated zeros
at $T = 1$ and $T=\rho\equiv e^{i \pi/6}$
in the fundamental domain $\cal D$ for $T$ (see. Fig. 1),
\goodbreak\midinsert{\hrf\vskip3.in\line{\noindent
Figure 1.~~~Fundamental
domain for $PSL(2,{\bf Z})$.\hfill}\hrf}\endinsert
{\it i.e.} when $|\partial_T
j(T)|^2=0 $\refmark{\modularform}\ .
Other isolated degenerate minima might as well arise when
$|\partial_j   P(j)|^2=0$.
Then, the mass per unit area of the domain wall can be written
as:\refmark{\vafa ,\AT }
$$
\mu = \int_{-\infty}^{\infty} dz\, G_{T \bar T}
| \partial_z  T - e^{i\theta} G^{T \bar T} \partial_{\bar T}
\bar W(\bar T)|^2  + 2 Re (e^{- i \theta} \Delta W)
\eqno (2)
$$
where
$\Delta W \equiv  W(T(z = \infty)) - W(T(z= -\infty))$. The arbitrary
phase
$\theta$ has to  be chosen such that
 $e^{i \theta} = \Delta W / |\Delta W|$,
thus maximizing the cross term in Eq. (2).
Then, we find $\mu \ge K \equiv 2 |\Delta W|$, where $K$ denotes the kink
number. Since $\partial_T W$
is analytic in $T$, the line
integral over $T$ is \sl path independent \rm
as for a conservative force. The minimum is obtained only if
the Bogomol'nyi bound $\partial_z T(z) = G^{T \bar T} e^{i \theta}
\partial_{\bar T} \bar W(\bar T(z))$
is saturated. In this case
$
\partial_z W(T(z)) = G^{T \bar T} e^{i \theta} |\partial_T W(T(z))|^2
$,
which implies that the phase of $\partial_z W$ does not change with $z$.
Thus, the supersymmetric domain wall is a mapping from the
 z-axis $[-\infty,
\infty]$ to a \sl straight line \rm  connecting between two degenerate
vacua in the $W$-plane.
We would like to emphasize that this result is general; it
applies to any globally supersymmetric theory with disconnected
degenerate minima that preserve supersymmetry.

For the superpotential, {\it e.g.}
$W (T) =  j(T)$
 the potential has two isolated  degenerate
minima at $T = 1$ and $T=\rho\equiv e^{i \pi/6}$ (see fig. 2 for
the potential along the geodesic $T=e^{(i\phi)}, \phi=\{ -\pi /6,
\pi /6\}$.
 At these fixed points,
$j(T=\rho) = 0$ and $j(T=1) = 1728$. Therefore, the mass per unit area
is $\mu = 2 \times 1728 $. The explicit solution for
$T=e^{(i\phi (z))}$ is displayed on Fig. 3.
\goodbreak\midinsert{
\setbox2=\vtop{\hsize=3.1in\noindent
Figure 2.~~~Global modular invariant potential along
the geodesic $T=e^{i\phi(z)}$.  Potential $V(\phi)$ plotted in
 units of $10^7$.}
\setbox3=\vtop{\hsize=3.1in\noindent
Figure 3.  Global domain wall $T(z)=e^{i\phi(z)}$
for modular invariant  potential. Length $z$ plotted in units of
$2\times10^{-5}$.}
\hrf\vskip4.in\line{\box2\hfill\box3}
\hrf}\endinsert
Other cases can be worked out analogously\refmark{\CGR}\ .
\endpage
\line{\bf 3. Local Stringy Domain Walls\hfill}
 \medskip

The case with  gravity
 restored
\foot{We use
the conventions: $\gamma^{\mu}=e^{\mu}_{a}\gamma^{a}$ where
$\gamma^{a}$ are the flat spacetime Dirac matrices satisfying
$\{\gamma^{a},\gamma^{b}\}=2\eta^{ab}$; $e^{a}_{\mu}e^{\mu}_{b}
= \delta^{a}_{b}$; $a=0,...3$; $\mu=t,x,y,z$; $\overline{\psi} =
\psi^{\dagger}\gamma^{t}$; $(+,-,-,-)$ space-time signature;
and dimensions such that
$8\pi G_N \equiv 1$.}
has a K\"ahler potential $K=-3 \log (T+\bar T)$ and the superpotential
should transform
as a weight $-3$
 modular function under modular
 transform\-a\-ti\-ons\refmark{\Ferrara ,\Cvetic} .
The simplest choice, with
supersymmetric vacua is with the superpotential
$$
W (T) = {\Omega (S)
  j(T) \over \eta (T)^6}.
\eqno (3)
$$
Here, $\eta(T)$ is the Dedekind eta function, a modular form of weight
$1/2$ and $j(T)$ is a modularly invariant function\refmark{
\modularform} .
The potential is of the following form:\foot{Note, that
the potential depends on a prefactor $\Omega (S)$ which depends
exponentially on the dilaton field $S$. Here we assume that
supersymmetry is not broken in the $S$  sector, \ie\
$D_SW=0$. On figures $\Omega =1 $ was used.
}

$$
V(T, \bar T) = \Omega (S)
{ 3 |j|^2 \over (T + \bar T)^3 |\eta|^{12}}
( |{(T + \bar T) \over 3} ({\partial_T j \over j} +
 {3 \over 2 \pi} \hat G_2) |^2 -1)
\eqno (4)
$$
where
$
\hat G_2=-{4\pi}\partial_T \eta/{\eta} -2\pi/(T+\bar T)$ is the
 Eisenstein
function of weight 2\refmark{\modularform} .
The scalar potential (4) has
two isolated supersymmetric
minima one at $T=1$ and one at $T=\rho$\refmark{\Cvetic} .
At these two supersymmetric minima, the superpotential takes values
$j(\rho)=0$ and $j(1)=1728$. This in turn implies that the
supersymmetric minima of the potential are non-degenerate.  At $T=1$ one
has an anti-deSitter space with cosmological constant
$-3|W(T=1)|^2e^{K(T=1)}$ and at $T=\rho $ the cosmological constant
is zero.  Even though the two supersymmetric minima of the matter
potential are not degenerate
(see Fig.~4 for the scalar potential along
the geodesic $T=e^{(i\phi)}, \phi=\{ -\pi /6,\pi /6\}$.
).  there does exist a stable domain wall solution interpolating
between them.
\goodbreak\midinsert{\hrf\vskip4.in {\noindent Figure 4.~~~Local modular
invariant potential along the geodesic $T(z)=e^{i\phi(z)}$. Potential
$V(\phi)$ plotted in units of $10^8$.}\hfill\break
\hrf}\endinsert
\medskip
\line{\bf 3.1 Minimal Energy Solution for Supersymmetric
 Stringy Domain Walls\hfill}
\medskip
 We now minimize the domain wall mass density.  Details of the
derivations are given in Ref. {\CGR}.
The analysis is general and can be applied to any supergravity
theory with isolated
supersymmetric vacua\refmark{\CGR}.
The bound
is a generalization of the global result.
We employ the results of
Ref.{\WITTEN}  and   Ref.{\NESTER}
which addressed the positivity of the ADM
mass in general relativity, as well as certain
generalizations to  anti-de Sitter backgrounds\refmark{\GHW}.

  By the planar
 symmetry, the most general {\it static}
 Ansatz  for  the metric
 in which the domain wall is oriented parallel to $(x,y)$ plane
is
$$ ds^2 = A( z ) (-dt^2 + d z^2) + B( z ) (dx^2 + dy^2). \eqno  (5)
$$

Consider the supersymmetry charge density
$$
Q[\epsilon] = \int_{\partial \Sigma} \bar{\epsilon}
\gamma^{\mu \nu \rho} \psi_{\rho} d\Sigma_{\mu \nu}
\eqno (6)$$
where $\epsilon$ is a commuting Majorana spinor,
$\psi_{\rho}$ is the spin $3/2$ gravitino field, and
$\Sigma$ is a spacelike hypersurface.
We take a supersymmetry variation of $Q[\epsilon]$ with respect to
another commuting Majorana spinor $\epsilon$
$$
\eqalign
{
\delta_{\epsilon} Q[\epsilon]& \equiv \{Q[\epsilon],
\bar{Q}[\epsilon]\}  \cr
&= \int_{\partial \Sigma}N^{\mu \nu} d\Sigma_{\mu \nu}
= 2\int_{\Sigma}\nabla_{\nu}N^{\mu \nu} d\Sigma_{\mu}  \cr
}
\eqno (7)$$
where
$N^{\mu \nu} = \bar \epsilon\gamma^{\mu \nu \rho}
\hat\nabla_{\rho} \epsilon $ is a generalized
Nester's form\refmark{\NESTER}.  Here
$\hat\nabla_{\rho}\epsilon \equiv
\delta_{\epsilon}\psi_{\rho} =
[2\nabla_{\rho} + ie^{K \over 2}(WP_{R} + \bar{W}P_{L})\gamma_{\rho}
 - Im(K_{T}\partial_{\rho}T)\gamma^{5}]\epsilon$ and
 $\nabla_{\mu}\epsilon = (\partial_{\mu}
  + {1\over2}\omega^{ab}_{\mu}\sigma_{ab})\epsilon$.
 In  (7)
the last equality follows
from Stoke's law.

We are concerned with supercharge \sl density \rm and thus
insist upon only $SO(1,1)$ covariance in the $z$ and $t$ directions.
This in turn implies
that the space-like
 hypersurface
$\Sigma$ in eq. (7)
is the $z-$axis
with measure
$d\Sigma_{\mu} = (d\Sigma_{t},0,0,0) =
|g_{tt}g_{zz}|^{1\over 2}dz$.  The boundary
$\partial \Sigma$ are then the
two asymptotic points $z\rightarrow \pm\infty$.
Technical details in obtaining the explicit form of
eq.(7)
are given in Ref.~\CGR.              Here we only quote the final
results.

  The volume integral
yields:
$$
  2\int_{\Sigma}\nabla_{\nu}N^{\mu \nu} d\Sigma_{\mu}=
\int_{-\infty}^{\infty}
[-\delta_{\epsilon}\psi^\dagger_i g^{ij} \delta_{\epsilon}\psi_j +
K_{T \bar T}\delta_{\epsilon}\chi^\dagger \delta_{\epsilon}\chi]dz
\ge0\eqno (8)          $$
where $\delta_{\epsilon}\psi_{i}$ and $\delta_{\epsilon}\chi$
are the supersymmetry variations of the fermionic fields in the
bosonic backgrounds.

Analysis of the surface
integral in (7)
yields two terms: $(1)$ The ADM mass density of configuration,
denoted $\sigma$ and
 $(2)$ The topological charge density,
denoted $C$ (see Ref.~\CGR).

Positivity of the volume integral
translates into the bound
$$
\sigma \ge
 |C|  $$
which is saturated iff $\delta_\epsilon Q[\epsilon]=0$. In this
case the bosonic backgrounds are supersymmetric; \ie\
they satisfy
$\delta\psi_{\mu} = 0$   and
$\delta\chi = 0$ (see eq.(8)).\foot{
$\delta_\epsilon Q[\epsilon] = 0$ seems to only require
$\delta_{\epsilon}\psi_{i} = 0$ with
 $i \ne t$.  However, in order for
$\delta_{\epsilon}\psi_{i} = 0$
for an \sl arbitrary \rm space-like hypersurface,
one in fact requires
$\delta_{\epsilon}\psi_{\mu} = 0$
for $\mu = t,x,y,z$ ~\refmark\WITTEN .}

The solution of self-dual equations yields:
$$
\eqalign{
\partial_z T(z) &= -\zeta \sqrt{A}|W|
e^{K\over 2}K^{T \bar T}
{D_{\overline{T}}\overline{W}\over \overline{W}},\cr
\partial_{z}lnA &=
\partial_{z}lnB =
2\zeta\sqrt{A}|W|e^{K\over 2}, \cr
Im&(\partial_{z}T{D_{T}W\over W}) = 0}
\eqno (9)$$
where $\zeta
=\pm1$ can change only at the point where $W$ vanishes.

We now comment on these three equations.

(i) The first equation in (9) is a local
generalization of the
global result. It is evident that $\partial_z T(z)
\rightarrow 0$ as one approaches the supersymmetric minima,
\ie\ $D_T W=0$, thus indicating
a domain wall configuration, however no constraint is put on the
degeneracy of vacua.
\goodbreak\midinsert{\setbox5=\vtop{\hsize=3.1in\noindent
Figure 5.~~~Local domain
wall $T(z)=e^{i\phi(z)}$ for modular invariant potential.  Length $z$
plotted in units of $4\times10^{-6}$.}
\setbox6=\vtop{\hsize=3.1in\noindent
Figure 6.~~~Metric function
$A(z)$ for modular invariant potential.  Length $z$ plotted in units
of $4\times10^{-6}$. }
\hrf\vskip4.in \line{\box5\hfill\box6}
\hrf}\endinsert
See Fig. 5 for the solution
$T=e^{ i\phi(z) }, \phi=\{ -\pi/6, 0\}$.

(ii)
The second equation  in (9), \ie\ the
 equation for the  metric, implies that we can always
rescale the space-time coordinates to bring $A = B$. Thus, our
metric Ansatz is reduced to a class of conformally flat metrics
with $z$-dependent conformal factor.
The asymptotic behaviour of the metric depends on whether
the supersymmetric vacuum is Minkowski ( $|W_{\pm \infty}|=0$) or
anti-deSitter ($|W_{\pm \infty}|\ne 0$).
In the first case the metric equation gives
$A\rightarrow const$,
while in the second case $A\rightarrow const'/z^2$,
which are the proper asymptotic behaviours in Minkowski
and anti-de Sitter space-times, respectively.
In the case of $W=j(T)\eta^{-6}(T)
$, at $T=e^{i\pi/6}$ the metric goes to a constant
($W=0$) and at $T=1$ the metric falls off at $1/|W(1)|^2e^{K(1)}z^2$.
See Fig. (6) for the explicit form of the metric $A(z)$.

(iii)
The third equation in  (9)
 describes a \sl geodesic path \rm
between two supersymmetric vacua
in the supergravity potential space $e^{K/2} W \in \bf C$ when
 mapped from the
$z$-axis $(-\infty, + \infty)$.
Here, we
 would like to contrast the geodesic equation in (9)
with the  geodesic  in the
global supersymmetric case. In the global case the
geodesics
are straight lines in the $W-$plane (see discussion after eq.
(2)).
On the other hand, the local
 geodesic equation in
 the limit $G_{N} \to 0$ (global
limit of the local supersymmetric theory)
leads to the geodesic equation
$Im({\partial_{z}W\over W})\equiv
\partial_z \vartheta = 0$
where $W$ has been written
as $W(z) =|W|e^{i\vartheta}$.
This in turn implies that
 as $G_N\rightarrow 0$ the geodesic equation
reduces to the constraint that  $W(z)$
has to be \sl a  straight line passing through the
origin\rm ;
\ie\ the phase of $W$ has to be
 \sl constant mod $\pi$ \rm. In the case of $W=j(T)\eta^{-6}(T)$
one can prove that the geodesic corresponds to $T=e^{i\phi(z)}$.
First note  $\partial_{T}j$ and $\hat G_2$ are both modular forms of
weight 2 while $j$ is the absolute modular invariant function.
The results $\partial_{T}j ({1\over T})=-T^2\partial_{T}j(T)$,
$\hat{G}_2 ({1\over T})=-T^2\hat{G}_2 (T)$, and $j(e^{i\phi})=
         j(e^{-i\phi})  $ imply $Im({{D_T W}\over W}\partial_z T)=0$
for $T=e^{i\phi(z)}$. Therefore
$T=e^{i\phi(z)}$ satisfies the geodesic equation.
Thus, the geodesic equation is the same as in the global case.

The energy density of the minimal energy solution can be written
as\refmark{\CGR}

$$
\sigma = |C| \equiv 2 |(\zeta|We^{K \over 2}|)_{z=+\infty}
-(\zeta|We^{K \over 2}|)_{z=-\infty}| = 2|\Delta (\zeta|W e^{K\over
2}|)|
\eqno (10)$$

Again in the case with $W$ as defined in (3),
 $\sigma=2W(T=1)e^{K(1)/2}$.
Eq. (10) constitutes a
generalization of the global case .

Comparing this local example with the corresponding
global supersymmetric modular invariant theory, both cases
are similar; e.g.,
the two  isolated supersymmetric minima
are at $T=\rho$ and $T=1$ and the geodesic is the same
in both cases. However, a significant difference is that
in the local case the minima are {\it
not degenerate}; \ie\ at $T=\rho$
the cosmological constant is zero, while at $T=1$ the
cosmological constant is negative.
In addition, these domain walls represent  a new class of
domain walls
beyond those classified in Ref. {\Ipser}; they are {\it static},
reflection asymmetric domain walls interpolating between non-degenerate
vacua.

The above     example is representative  of a situation
where the study of  a global supersymmetric domain wall
is readily generalizable to a   local supersymmetric theory.
One may
be tempted to conclude that all the supersymmetric domain walls in
the global supersymmetric theory automatically remain as supersymmetric
domain wall solutions even after gravity is turned on.
However, this is not always the case.

Consider another modular invariant superpotential:
$$
W  = j(T) (j(T) - 1728)
\eqno (11)
$$
There are three isolated
global supersymmetric minima at
$T=1, \rho$ and $\partial W/\partial j = 2j(T)-1728 =0$.
Therefore, we expect two  domain walls
interpolating between each of the
two adjacent vacua.
In the supergravity case we find the minima
$T=1$ and $T=\rho$ remain supersymmetric minima. They both
have zero cosmological
constant since the
superpotential
 vanishes at these two points.\foot{ Note, again, that
$j(\rho)=0$ and $j(1)=1728$.}
Additionally, there
 is a local minimum with
positive cosmological constant at $T_{3}$ which is
in the neighborhood of the point
$j^{-1}(864) \in \cal D$.  However,
this point
is not supersymmetric since
$D_T W =[\partial_{T}W+{3\over{2\pi}}\hat G_2 W]|_{T=T_3}\ne 0$.
Thus, the domain wall  interpolating between $T=1$ and  $T_3$
(or between $T_3$ and $T= e^{i\pi/6}$) is not stable
since the minimum at $T_3$
is a non-supersymmetric \sl de-Sitter \rm minimum.
Also, the
wall interpolating directly
 between the supersymmetric vacua at $T=1$ and $T=e^{i\pi/6}$
does not exist either as
the superpotential vanishes at these vacua and thus there
is no energy associated with such a wall.

\medskip
\line{\bf 3.2 Non-perturbative Stability of Supersymmetric
String Vacua \hfill}
\medskip

The analysis  of the above local supersymmetric domain wall
solution interpolates between
two non-degenerate
vacua of the supergravity matter potential, \eg\
one with zero and another with negative
cosmological constant.
The existence of such
static
domain walls has strong implications for the
non-perturbative stability of supersymmetric vacua, and thus also
for supersymmetric supersting vacua\refmark{\CGRII}\ .
The obtained result
is intimately related to the $O(4)$ symmetric
bubbles of the false vacuum decay in the presence of
gravity\refmark{\COLEDEL} .
In Ref.{\COLEDEL}
Coleman and DeLuccia found
that a false vacuum decay from the Minkowski space-time
to anti-deSitter space-time  cannot take place
\sl unless \rm the matter vacuum energy difference
$\epsilon =-V(true)$
meets an inequality
$$
\epsilon \ge {3 \over 4 } \sigma^2
\eqno (12)
$$
in which $\sigma$ denotes the energy density stored in the bubble wall.
In the case of false vacuum decay from anti-deSitter to anti-deSitter
space-time one arrives\refmark{\CGRII} at the same equation
(12) with
$\epsilon \equiv (\sqrt{-V(true)}-\sqrt{-V(false)})^2$

The residual energy after materializing the bubble wall
accelerates the wall asymptotically to the speed of light.
Also as the energy difference $\epsilon$ approaches the minimum of
the Coleman-DeLuccia
bound
(12) the radius of the $O(4)$ invariant bubble wall becomes
indefinitely large. Precisely at the saturation limit,
$$\sigma = \sigma_c \equiv 2 \sqrt {\epsilon \over 3}. \eqno (13)$$
No kinetic energy
is available for the wall to accelerate to the speed of light, and the
wall radius becomes infinite, \ie\ becomes planar.
The resulting configuration of the $O(4)$ bubble is a
time-independent and infinite planar domain wall dividing
the Minkowski space-time from the anti-deSitter space-time.
In the
supergravity theory, $\sigma_c = 2 \sqrt {\epsilon \over 3}
= 2 e^{K/2} |W(true)|$ which coincides with the topological charge
$|C|
= 2 |\Delta (\zeta e^{K/2} |W|)|$ (see eq. (10)).
 Thus, the critical Coleman-DeLuccia
bubble
wall in supergravity theory saturates the Bogomol'nyi bound,
and hence,
this is a special class of the supersymmetric domain wall
described above.
This result has strong implications for the stability of supersymmetric
vacua in general, and superstring vacua in particular; namely,
supersymmetric vacua are non-perturbatively stable against false
vacuum decay \refmark{\CGRII}. This result completes a perturbative
analysis  $\kappa=8\pi G_N
\to 0$
of Weinberg \refmark{\weinberg}.

As a collorary, non-supersymmetric vacua are unstable
against false vacuum decay because in this case the
Coleman-DeLuccia bound (12)
can be satisfied; note, that in the
non-supersymmetric vacuum satisfies the inequality
$V>-3|W|^2e^K$.
Thus, a non-supersymmetric
vacuum  would  decay into a
supersymmetric one    if there
exists a supersymmetric one.
\medskip
\line{\bf 3.3 Non-Supersymmetric Stringy Domain Walls\hfill}
 \medskip
In view of non-perturbative  stability of supersymmetric vacua
one is compelled to search for non-perturbatively induced potentials
 for the modulus fields $T$ with  only non-supersymmetric
vacua; in  this case the non-supersymmetric vacuum cannot
decay into existing supersymmetric one.

It turns out that
the simplest form of
 such a superpotential is\refmark{\Ferrara ,\Font} :
$$ W=\Omega (S) \eta ^{-6}(T) \eqno (14)$$
Within the
fundamental domain $\cal D$ (see Fig.1) the corresponding
potential has
only one minimum   at $T=1.2$\refmark{\Font ,\Cvetic}
which also breaks supersymmetry.  In addition, the
superpotential (14)  can be derived explicitly\refmark{\Font , \LM}
as an effective term due to  gaugino  condensation of the hidden
$E_8$
gauge group in  orbifold compactifications. It is thus
the best motivated   non-perturbatively
induced  superpotential in a class of superstring vacua.

 The  underlying $PSL(2,{\bf Z})$ symmetry of the theory
 implies \refmark{\Giveon ,\FONTII ,\CV
  }
 that there should be domain wall solutions
interpolating between such degenerate vacua of different
fundamental domains.

The nature and existence of such domain walls has
recently been studied in Ref.{\CD}. There are two classes
of domain walls associated with the superpotential (14).
The first class are
domain walls is associated with the  symmetry  transformation
$T\to T+i$, \ie\   the
discrete Peccei-Quinn symmetry. Thus the domain walls
interpolates between minima with
$T=1.2 $ and $T=1.2+i$.

 The nature of this domain wall is
closely related to the domain wall that exists for the  QCD  induced
potential of the Peccei-Quinn axion $\theta$
with only one
quark flavor, \ie\ $V=1-\cos\theta$. In this case
there is a domain wall interpolating between $\theta =0$ and
$\theta=2\pi$. The  domain wall
 is bounded by an axionic string which emerged
at the first stage of symmetry breaking of the global $U(1)$
Peccei-Quinn symmetry.
Analogously, in our case the role of the
 stringy axion field is played by the
imaginary part
of the $T$ field;  the domain wall interpolates
between $T=1.2$ and $T=1.2 + i$ and it is
 bound by stringy cosmic strings\refmark{\VAFA} of the type
$j(T(x+iy))= a(x+iy)+b$. Here, $x,y$ are the spatial
coordinates, a and b are arbitrary constants, and $j$ is
modularly invariant function.
The existence of these stringy cosmic strings
 is associated with the breakdown of the global non-compact
symmetry $SL(2,{\bf R})$ which is there in the  $T$ sector
to all orders in string loops, and is only broken by the
non-perturbative effects, like gaugino condensation.
Stringy cosmic strings have a natural scale\refmark{\VAFA}
 of ${\cal{O}}(1/\sqrt{\alpha '})
={\cal{O}} (10^{17}\hbox{GeV})$.

The second type of domain walls are        associated with
$T\to 1/T$, \ie\ the generator of the
non-compact symmetry  transformation of $PSL(2,{\bf Z})$. This
domain wall interpolates between
the minimum at $T=1.2$ and $T=1/1.2$. It is
   analogous to the
domain walls associated with $Z_2$ symmetry.
It is at first puzzling that
there would be such a domain wall, after all,
the points in the $T$ plane are related by the $T\to 1/T$
symmetry. However,
 points associated with $T\to
 1/T$ transformation
 {\it can} be probed
since they correspond to a different theory (with heavy
winding modes becoming light and vice versa) which happens
to be equivalent to the original theory.

There are   difficulties associated with the
 cosmological implications of the above domain walls.
First, the scale of the
domain wall depends on the scale $\Omega (S)$, \ie\ the scale
 at which     gaugino condensation
takes place.
This scale could be as low as ${\cal{O}}$(TeV)  or  as high
as ${\cal{O}}(10^{16}\hbox{GeV})$. The latter one is more plausible,
at least in the scenario where the hidden gauge group is $E_8$.
Thus, domain walls have to decay rapidly in order to  be consistent
with observations. One obvious mechanism would be
by invoking inflation. The second possibility is the
decay via choping by the stringy cosmic strings, as
long as the energy
scales of the two types of topological defects
(the domain walls and  the
strings) are  not too far apart.

Another difficulty  with the above scenario
is that
the Kibble mechanism for generating
stringy cosmic strings has not been established, yet.
Further study is of potential cosmological implications
of non-supersymmetric domain walls is
in progress\refmark{\CD} .

\medskip
{\bf 4. Other Topological Defects\hfill}
\medskip
We would now like to point out\refmark{\CQRprep}
 the existence of other global
topological defects, like global monopole-type
 and texture-type defects in the moduli sector with more than one
 modulus.
  We shall illustrate the idea using
examples based on the so called flat backgrounds, {\it i.e.}
generalization of $SL(2,{\bf R})$.

For that purpose we shall study the simplest example of $Z_4$ manifold
with continuous symmetry $SU(2,2)$ on the four moduli
$$
\bf T\equiv\left[\matrix{
T_{11} & T_{12} \cr
T_{21} & T_{22} \cr}\right]\eqno(14) $$
of compactified space.  Note that the moduli $\bf T$ live on the coset
$SU(2,2)/SU(2)\times SU(2)\times U(1)$.  The continuous non-compact
symmetry $SU(2,2)$ is an {\it exact}
symmetry\refmark{\CLO} at least at the
string tree-level.  Note that this continuous symmetry in the modulus
could be broken down to the discrete subgroup $SU(2,2,Z)$ due to
nonperturbative effects, {\it e.g.} gaugino condensation and/or axionic
instanton effects.  At this point we shall assume that this
non-compact continuous
symmetry is preserved in the modulus sector
all the way to low energies  and is not
broken by non-perturbative effects.
However, one should keep in mind that $SU(2,2,Z)$ is
the vacuum symmetry and thus the $\bf T$ fields should live in the
fundamental domain of $SU(2,2,Z)$.

The maximal compact symmetry of $SU(2,2)$ is $SU(2)_A\times
SU(2)_B\times U(1)\subset SU(2)_{A+B}$.  Note also that in projective
coordinates\refmark{\Gilm}\ :
$  \bf Z=(1-\bf T)/(1+\bf T)
.$
$\bf Z$ transforms as
${\bf 1}+{\bf 3}$ under $SU(2)_{A+B}$.  The Ansatz
${\bf Z}=\sum_{a=1}^3\sigma_a V_a$
with $V_a=f(r)x_a/r$ ensures the map of
$\bf Z$ on the $S^2$.

The $\bf Z$ fields have {\it no}
potential to all orders in string loops.  Thus the kinetic energy
term\refmark{\CLO}\
shrinks $f\rightarrow0$ due to Derrick's theorem, however it
could be stabilized by higher derivative terms.
Such higher derivative terms
arise  even at the tree level of the
string theory.  They should respect the
noncompact $SU(2,2)$ symmetry.  Also, if one sticks to terms with at most
two time derivatives, one has a unique form for the terms that involve
four derivatives,
which is very similar
in nature to the Skyrme term\refmark{\Skyr} in the
Skyrmion model and can serve the same role as the stabilizing term.
The energy stored in such a configuration is finite and is governed
by the scale of $\alpha '$.
This is different from the standard global monopole
configuration\refmark{\Vilenkin}\ ,
which has linearly divergent energy and thus long
range interaction relevant for large scale formation.

Texture-type configurations,
can also occur within
this sector.  Namely, the $\bf Z$
fields transform as $\bf 4$ under the
compact symmetry $SU(2)_A\times SU(2)_B\sim SO(4)$ and thus the Ansatz:
${\bf Z}=a(r)+b(r)\sum_{a=1}^3\sigma_a x_a/r$
is mapped onto $S^3$.

The above
studied  configurations are much milder defects than strings
and domain walls and they  have
finite range and thus finite energy.  Further study and
 cosmological implications of such global defects is necessary.

I would like to thank my collaborators R.~Davis,
S.~Griffies, F.~Quevedo, and
S.-J.~Rey, for many fruitful discussions and enjoyable collaborations.
I would also like to thank the Aspen Center for Physics, the
International Centre for Theoretical Physics, Trieste, and CERN for
their hospitality.  The work is supported in part by the U.S. DOE Grant
DE--22418--281 and by a grant from
University of Pennsylvania Research Foundation and by the NATO
Research Grant \#900--700.
\refout\end